\documentclass{sig-alternate}
\usepackage{epsfig}

\makeatletter
\def\@copyrightspace{\relax}
\makeatother

\begin{document}


\title{Visual Concept Ontology for Image Annotations}

\numberofauthors{3}
\author{%
\alignauthor
Jan Botorek\\
       \affaddr{Faculty of Informatics}\\
       \affaddr{Masaryk University}\\
       \affaddr{Brno, Czech Republic}\\
       \email{xbotorek@fi.muni.cz}
\alignauthor
Petra Budikova\\
       \affaddr{Faculty of Informatics}\\
       \affaddr{Masaryk University}\\
       \affaddr{Brno, Czech Republic}\\
       \email{budikova@fi.muni.cz}
\alignauthor
Pavel Zezula\\
       \affaddr{Faculty of Informatics}\\
       \affaddr{Masaryk University}\\
       \affaddr{Brno, Czech Republic}\\
       \email{zezula@fi.muni.cz}
}

\maketitle

\begin{abstract}
In spite of the development of content-based data management, text-based searching
remains the primary means of multimedia retrieval in many areas. Automatic creation
of text metadata is thus a crucial tool for increasing the findability of multimedia objects.
Search-based annotation tools try to provide content-descriptive keywords
by exploiting web data, which are easily available but unstructured and noisy.
Such data need to be analyzed with the help of semantic resources
that provide knowledge about objects and relationships in a given domain.
In this paper, we focus on the task of general-purpose image annotation
and present the VCO, a new ontology of visual concepts developed as a part of image
annotation framework. The ontology is linked with the WordNet lexical database,
so the annotation tools can easily integrate
information from both these resources.


\end{abstract}




\section{Introduction}
\label{sec:Intro}

As more and more information becomes expressed by images or videos, effective
management of multimedia data remains one of the priorities in information processing.
In spite of the development of content-based retrieval, text-based searching continues
to be the most frequent data access method.
Consequently, multimedia objects need to be accompanied by content-describing text metadata to be findable.
Unfortunately, the descriptions are often of uncertain quality or not available at all,
since their manual creation is a tedious and time-consuming task.
In this situation, automatic annotation of multimedia content and text metadata refinement methods
are highly desirable.

\begin{figure}[t]
\centerline{
\includegraphics[width=0.47\textwidth]{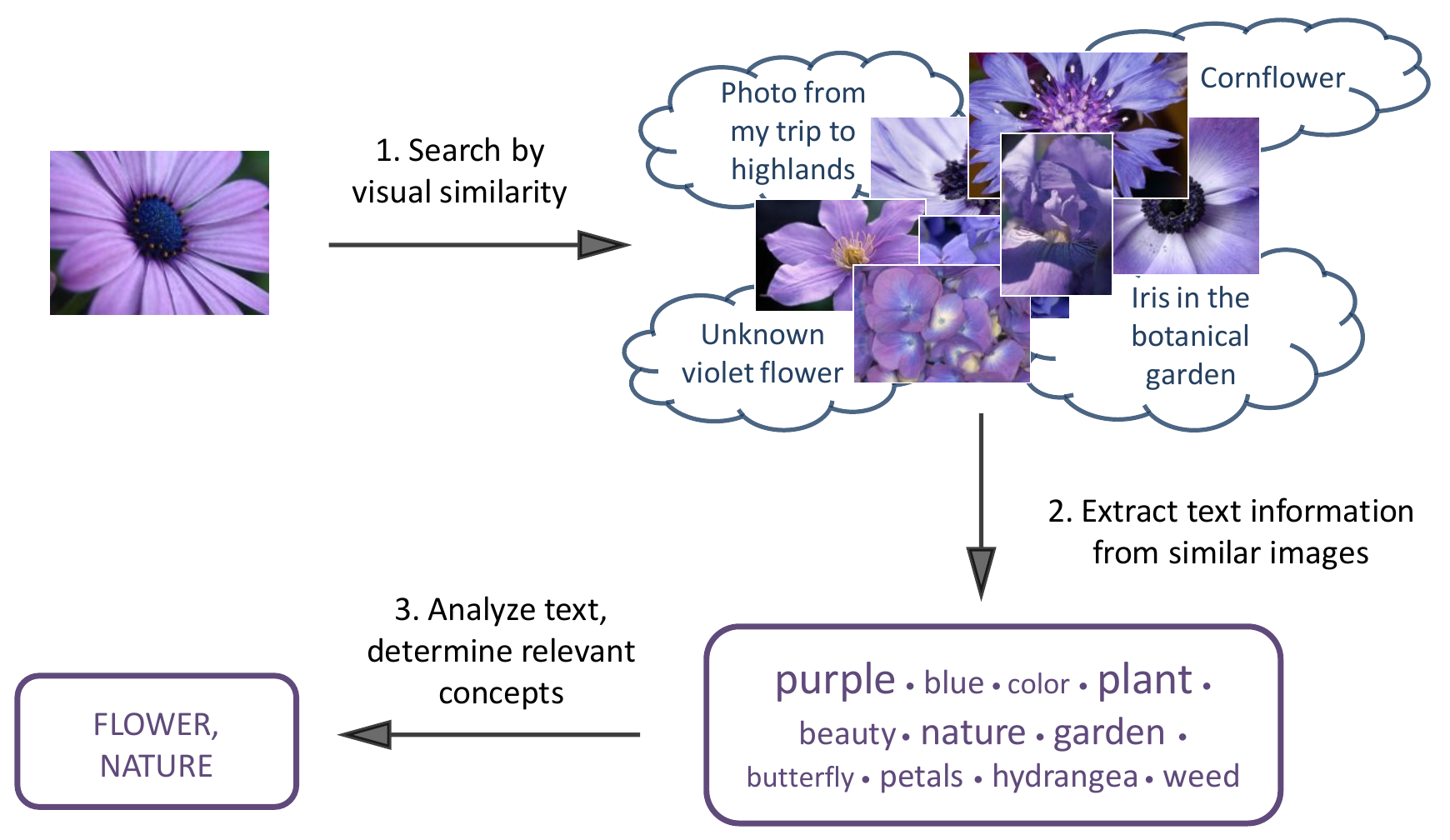}
}
\caption{\label{fig:annotIllustration}Search-based image annotation scheme.
}
\end{figure}

In this paper, we focus on automatic annotation of images. We assume that
the annotation task is defined by a query image $I$ and a vocabulary $V$ of candidate concepts,
and the annotation function $f_A$ assigns to each concept $c \in V$ its probability of being relevant for $I$.
Traditional {\em model-based approaches} use manually-labeled training data to create
classifiers for all concepts in $V$, which are then used to determine the relevance of any concept for image $I$.
Such techniques are known to perform well in narrow-domain
tasks (e.g. medical image classification), but are limited by the availability of high-quality
training data, which may become a significant bottleneck when the vocabulary $V$ is large.
Therefore, an alternative {\em search-based annotation} has been proposed recently for general-purpose annotations.
This method tries to exploit noisy but voluminous
data that is easily available e.g. on the web. Specifically, content-based retrieval is applied
to obtain images that are visually similar to $I$, and their metadata is used for reasoning about the
probability of concepts from $V$~\cite{BatkoBBZ13,ZhangIL12}. The whole process is schematically depicted in Figure~\ref{fig:annotIllustration}.

%
%

The search-based approach allows us to take advantage of the digital data explosion, but also poses multiple challenges that need
to be solved to achieve reasonable annotation quality. One of the crucial problems is determining the relevance of candidate
concepts using the metadata of similar images (i.e. step 3 in Figure~\ref{fig:annotIllustration}). It may be helpful to consider a
parallel -- let us imagine a person trying to guess what is depicted in a hidden image, using only visually similar images selected by a computer.
The similar images may be noisy in terms of semantic content (due to the well-known {\em semantic gap} problem, an orange ball
is likely to be evaluated as similar to both an orange fruit and the sun),
the guessing person thus needs to look for connected topics and try to infer the most probable content of the hidden image. When
we replace the person by some automated system, the task becomes even more difficult. The system does not understand
an image per se and needs to work with its metadata, which again are likely to be noisy and further complicated by natural language properties
such as ambiguity, use of synonyms, abstraction, etc. But most importantly, the computer lacks human experience with
real world that allows us to decide which topics are related and infer further information.

To be able to make sense of the similar images' metadata, the annotation system thus needs two types of knowledge sources:
1) computer-processable information about linguistic relations between words, and
2) ontologies that describe the semantics of real-world concepts and links between them.
The first resource can be considered domain-independent, but the latter should
be tailored for a particular domain to reflect the importance of individual concepts
and their relations in the given context.
For general-purpose image annotation, the domain of interest consists of so-called
{\em visual concepts}, i.e. objects or abstract notions that are typically depicted in photos.
Although some existing ontologies cover this domain, none of them entirely suits the needs of the annotation task~\cite{TouschHA12}.
Specifically, these ontologies are either too complex or unsuitably
structured, so that some important semantic connections between visual concepts are missing.

Therefore, we propose a new {\em Visual Concept Ontology (VCO)}, which organizes common visual
concepts that appear in image annotations.
The VCO is based on the Word\-Net lexical
database, a linguistic tool frequently used in information mining.
By encapsulating relevant parts of the
Word\-Net hierarchy into a more concise structure,
the VCO becomes a strong tool for analysis of image descriptions.
\section{Existing Semantic Resources}
\label{sec:RelWork}

Before introducing the VCO, let us briefly survey existing semantic resources and discuss their limitations with
respect to the image annotation task.
We focus on ontologies and tools that can be used for processing of general-purpose images (such as those found
in personal photo collections), leaving out domain-specific resources for medicine, arts, etc.

\paragraph*{WordNet}
The WordNet~\cite{WordNet} is a comprehensive semantic tool interlinking dictionary, thesaurus and language grammar
book.
The basic building block of WordNet hierarchy is a {\em synset},
an object which unifies synonymous words into a single item. 
On top of synsets, different semantic relations are encoded in the WordNet structure, e.g.
hypernymy/hyponymy (super-type and sub-type relation) or meronymy (part-whole relation).
The WordNet structure is divided into four
independent categories according to the parts of speech encoded -- nouns, verbs, adjectives, and adverbs.
The noun hierarchy, which is best developed, forms a tree structure corresponding to the hypernymy and hyponymy
relations among synsets.
In the current version of WordNet,
there are more than 82,000 synsets for 117,798 unique nouns stored in the noun hypernymy tree structure.

In context of search-based image annotation, the WordNet is very useful for identifying synonymous
words and discovering relationships between concepts~\cite{BatkoBBZ13,ZhangIL12}. However, the WordNet relationships only cover
linguistic dependencies, which are not satisfactory for image metadata analysis -- for instance,
there is no relationship linking ``roof'' and ``house'' although these words are clearly semantically related.
Another problem of the WordNet structure is uneven depth of individual branches of the noun hierarchy, which makes it difficult
to quantify the relatedness of any two synsets.

\paragraph*{ImageNet}
A well-known resource for image data processing is the ImageNet~\cite{ImageNet}, a database of images organized according to the WordNet hierarchy.
The ImageNet project aims at illustrating each WordNet synset by several hundred images, using a crouwdsourcing platform
to supervise the selection of images.
Currently, the database contains about 14 million images for nearly 22 000 noun synsets.

The ImageNet has been successfully engaged in many image processing tasks, but its applicability for
general-purpose annotations is limited. The database can be used as the reference
dataset, from which the similar images are retrieved; when sufficiently similar images are
found for the input image $I$, the annotation will be very precise. However, more complex scenes with multiple topics are not likely to be
found among ImageNet objects and therefore will not be annotated correctly. Less organized but significantly larger
collections of web images are therefore more
promising for the general-purpose annotations.


\paragraph*{Ontologies}
Ontologies are a standard tool for describing knowledge about concepts from a given domain.
In contrast to narrow-domain ontologies, which are created by a few domain experts,
broad-domain knowledge bases are often developed by cooperation of large user groups
(DBpedia\footnote{http://dbpedia.org}, Freebase\footnote{http://www.freebase.com/}, etc.).
The crowd-sourcing paradigm can facilitate wide coverage, but it is problematic to keep
the overall philosophy of the resource uniform, which complicates any utilization
of such ontology in automatic processing.
Therefore, some recent ontologies have been developed on top of existing, well-structured resources, especially the WordNet.
The YAGO ontology~\cite{YAGO} has been automatically created by aligning the WordNet with Wikipedia facts,
whereas the existing Kyoto ontology~\cite{Kyoto} has been semi-automatically mapped to the WordNet.
However, all these resources were designed primarily for text processing, which has different requirements
on the selection and categorization of concepts than analysis of image annotations.

A few attempts have also been made to establish an ontology for visual information. The LSCOM ontology~\cite{LSCOM}
specializes on concepts appearing in video news, and a simple ``Photo Tagging Ontology'' covering
100 concepts was issued with the ImageCLEF annotation task~\cite{Nowak11}. However, the first one specializes
only on the most frequent video news topics, whereas the second is too shallow to be applicable outside the ImageCLEF competitions.


%



\section{Visual Concept Ontology}
\label{sec:VCO}

\begin{figure*}[t!]
\centerline{
\includegraphics[width=\textwidth]{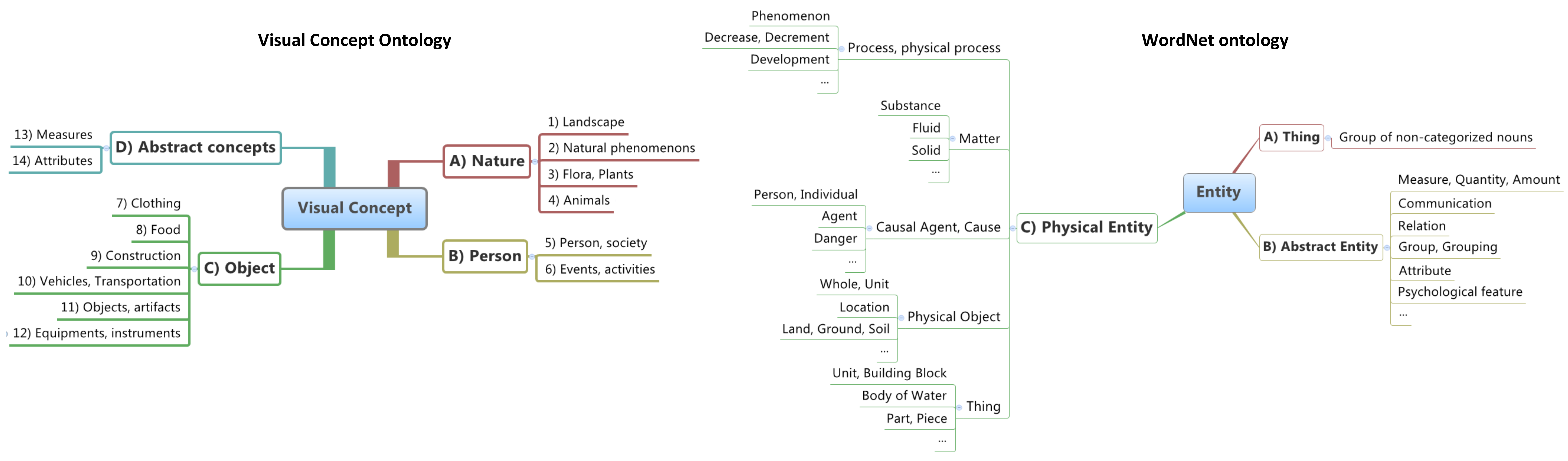}
}
\caption{\label{fig:ontologies}Comparison of VCO and WordNet top-level concepts.
}
\end{figure*}

As we discussed in the introduction, knowledge bases that provide information about visual concepts and their
relationships are much needed for search-based annotation. During the development of a general-purpose annotation system~\cite{BatkoBBZ13},
we discovered that none of the existing resources fulfills the specific needs of such application. Therefore, a new Visual Concept Ontology
was created for the specific needs of image content description.

The Visual Concept Ontology organizes frequent visual concepts into a hierarchical
structure and maintains semantic links between them.
Individual concepts are not defined formally in the VCO, but are specified by links to related WordNet
synsets.
The VCO taxonomy simplifies and flattens the WordNet hierarchy, removing concepts not relevant
to the visual domain and adding semantical connections between interrelated WordNet subtrees.
At the moment, VCO only covers nouns from the WordNet structure, as these are most frequently
used in image descriptions and have the strongest support in the WordNet.
The ontology is encoded by the Web Ontology Language OWL, a state-of-the-art formalism in knowledge
representation. Individual VCO concepts are represented by OWL classes, whereas the related WordNet synsets are modeled
as OWL individuals. 

\subsection{Selection of Concepts}
The objective of the VCO is to identify basic semantic categories of objects occurring in a picture
and organize them in a hierarchical structure. We are not interested in very specific categories
such as ``poodle dog'', which are well organized in the WordNet, but rather in determining reasonable
top-level categories of visual objects.
To choose suitable concepts for populating the VCO taxonomy, we utilized a combination of top-down and
bottom-up analysis of the WordNet hierarchy. The ontology construction process included the following four phases:

\vskip 0.6\baselineskip
\noindent
{\em 1. Significant synset extraction } In the first phase, we extracted significant synsets from the
WordNet noun hypernymy tree. For each synset, the number of all hyponyms was computed. If the number
was higher than a specified constant (experimentally set to 300), the synset was declared to be significant
and was included into a group of VCO concept candidates.
In this phase, more than 400 candidate synsets were identified.

\vskip 0.6\baselineskip
\noindent
{\em 2. General concept removal } In the next step, we manually examined the candidates and removed
very general synsets such as ``entity'' or ``thing''. Similar to stopwords in natural language, these
general concepts would be of little use in image understanding tasks. Therefore, we checked individual branches
of the border candidates tree and selected synsets that were specific enough to be included in the ontology.

This cleaning needed to be done manually, since individual branches of the WordNet noun tree significantly differ in depth.
To reach the ``white rhinoceros'' synset from the ``entity'' root synset, it is necessary to traverse 15 more
general synsets, but only 7 synsets separate the ``river Thames'' from the root. Therefore,
it was not possible to simply remove the first $k$ levels of the WordNet hierarchy.

In this phase, we also removed most descendants of the ``abstraction'' synset that are highly unlikely to
be visually represented in images (e.g. ``communication'' synset with descendants like ``Soprano'' or ``poetry'', or
``grouping'' synset covering concepts of ``overpopulation'' or ``baby-boom'').
However, several abstract concepts that can be meaningfully applied to images were retained, especially those regarding
time-related concepts (``Christmas'',``evening'') and human emotions (``joy'', ``sadness'').

\vskip 0.6\baselineskip
\noindent
{\em 3. Organization of classes } Having selected significant synsets with visually-relevant
content, we were able to establish the top-level VCO classes and individual sub-trees. The remaining
candidate synsets were explored and semantically similar ones were merged a joint class. In this
step, we thus established links between semantically interrelated synsets placed in different parts of the WordNet
hierarchy -- e.g. the ``roof'' and ``house'' synsets that are far apart in the WordNet.

\vskip 0.6\baselineskip
\noindent
{\em 4. Confrontation with existing ontologies } In the last phase, we examined other broad-domain ontologies to
check whether some important concepts were not missed. We particularly focused on inspection of basic
classes and overall architecture of the taxonomies. Specifically, YAGO~\cite{YAGO}, Sumo~\cite{NilesP01}, LSCOM~\cite{LSCOM} and the ImageCLEF
Photo Tagging Ontology~\cite{Nowak11} were utilized to check and fine-tune the VCO taxonomy.

\vskip 0.6\baselineskip
As a result of these four steps, we obtained 14 top-level classes that were further divided into 90 more specific sub-classes.
On top of these, a final high-level generalization was performed, producing 4 super-classes: {\em nature, person, object} and {\em abstract concepts}. The resulting hierarchy is outlined in Figure~\ref{fig:ontologies}; in contrast to the WordNet, the VCO taxonomy
is semantic rather than lexicographic and the top-level categories are suitable for human-readable labeling of image content.
Each VCO class is linked to one or several WordNet synsets, which define its semantics and allow to access a detailed hierarchy
of synsets relevant for the given topic.



\subsection{Relationships}
\label{sec:Relationships}

Relationships between ontology classes are a vital tool for describing semantics. In the VCO, we
define two basic types of relationships -- {\em class-to-class} and {\em class-to-individual}. The class-to-class
relationships describe semantic links between individual VCO concepts, especially
the standard sub/super-class relation between classes. 
Class-to-individual relationships connect VCO classes with {\em individuals} -- OWL objects
that represent related WordNet synsets. Inspired by the Kyoto ontology which also models
synset-to-concept links, we defined the following two types of class-to-individual relationship:
{\em equivalenceOf} link connects an individual to a class that is semantically equal to the synset
referenced by the individual, {\em superClassOf} links classes with individuals that are not
semantically equal to it but semantically belong to the given class.



%

\section{Possible Applications}
\label{sec:Applications}

\subsection{Hierarchical Image Annotation}

As we have discussed in the introduction, search-based annotation techniques are primarily intended
for tasks with wide annotation vocabularies.
Specifically, we aim at developing a general annotation tool that would work with the full English vocabulary
and provide keyword annotation with user-selected level of detail. The overall architecture of this tool is described
in~\cite{BatkoBBZ13}. The VCO ontology will be used for two purposes within the tool. First, it will be exploited during analysis
of the descriptions of similar images, as described earlier. Complementing the WordNet relationships with VCO categorization
will provide a richer set of links between concepts, and the VCO structure will also be used to prevent too
extensive searching for connections.
Second, after choosing the most probable
VCO and WordNet subtrees, the final annotation will be presented to the user using the VCO taxonomy, so that the user will be
able to easily access descriptive keywords on different levels of detail.


%

\subsection{Ground Truth Construction}

\begin{figure}[t!]
\centerline{
\includegraphics[width=0.47\textwidth]{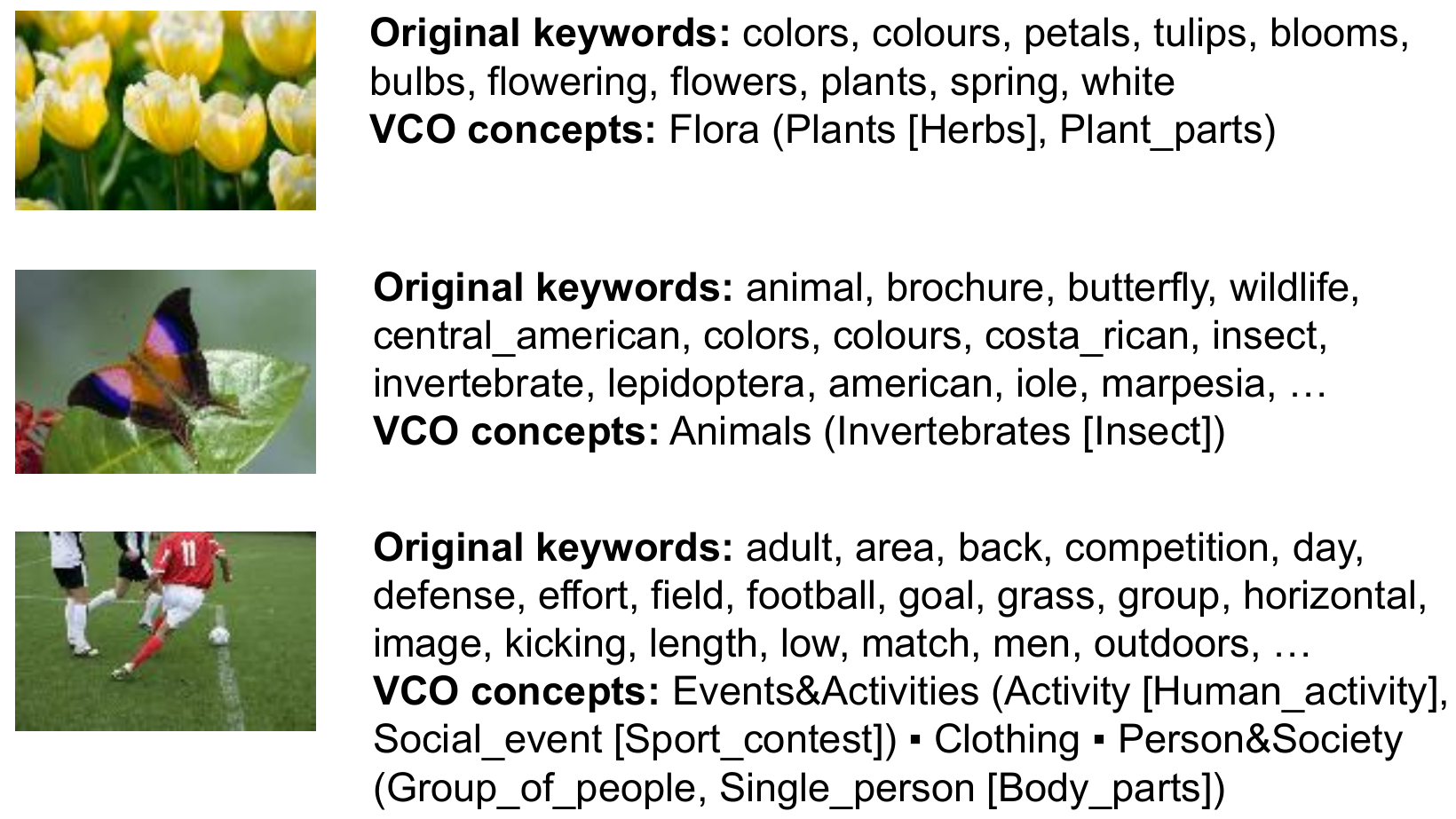}
}
\caption{\label{fig:evaluation}Author keywords vs. covering concepts.
}
\end{figure}

As a second possible application, we would like to discuss the problem of ground truth construction for the evaluation
of annotation techniques. Establishing evaluation data is known to be a difficult task, which is typically solved
by manual effort of a group of experts or using the crowdsourcing platforms. However, the former approach can
only generate small test collections, whereas the latter is likely to provide uncomplete or erroneous ground truth.
Using the VCO, we believe that the ground truth construction could be facilitated in a semi-automatic way,
removing much of the manual effort and guaranteeing a uniform vocabulary. Specifically, we suggest to take
some dataset of richly annotated images, e.g. the Profiset collection\footnote{http://disa.fi.muni.cz/results/software/profiset-testbed/}
of web-stock images, analyze the annotations using WordNet relationships, and replace them by the appropriate top-level categories and subcategories
from VCO. If the original annotations are rich and precise enough, which is typically true for stock data, there is a high chance that the automatically
selected categories will provide precise annotation of the image content. Any imprecisions can be removed
by a quick manual checking of the classification results. In contrast to the original annotations,
which are not applicable as a ground truth due to the unsupervised vocabulary, the new metadata will be selected from a fixed vocabulary
with a given level of detail. Figure~\ref{fig:evaluation} shows the transformed metadata of three random pictures from
the Profiset collection.

\section{Conclusions}
\label{sec:Conclusions}

This paper introduces the VCO ontology, which was created as a part of search-based image annotation framework.
The VCO provides a mapping between fundamental visual concepts and the WordNet
hierarchy, thus creating a useful connection between the image domain semantics and natural language.
The ontology can be used to analyze image descriptions and assist in categorization of image content,
but also to automatically extract category-based ground truth from web images with free-text annotations.
The complete ontology is available at
{\tt http://disa.fi.muni.cz/vco}.
\section*{Acknowledgments}
{
This work was supported by the Czech national research project GBP103/12/G084.
}

\bibliographystyle{abbrv}
\bibliography{ontologies}

\begin{thebibliography}{10}

\bibitem{BatkoBBZ13}
M.~Batko, J.~Botorek, P.~Bud\'{\i}kov{\'a}, and P.~Zezula.
\newblock {Content-Based Annotation and Classification Framework: A General
  Multi-Purpose Approach}.
\newblock In {\em IDEAS 2013}, pages 58--67. ACM, 2013.

\bibitem{ImageNet}
J.~Deng, W.~Dong, R.~Socher, L.-J. Li, K.~Li, and F.-F. Li.
\newblock {ImageNet: A large-scale hierarchical image database}.
\newblock In {\em CVPR}, pages 248--255. IEEE, 2009.

\bibitem{WordNet}
C.~Fellbaum, editor.
\newblock {\em {WordNet: An Electronic Lexical Database}}.
\newblock The MIT Press, 1998.

\bibitem{Kyoto}
E.~Laparra, G.~Rigau, and P.~Vossen.
\newblock {Mapping WordNet to the Kyoto ontology}.
\newblock In {\em LREC-2012}, pages 2584--2589, 2012.

\bibitem{LSCOM}
M.~R. Naphade, J.~R. Smith, J.~Tesic, S.-F. Chang, W.~H. Hsu, L.~S. Kennedy,
  A.~G. Hauptmann, and J.~Curtis.
\newblock Large-scale concept ontology for multimedia.
\newblock {\em IEEE MultiMedia}, 13(3):86--91, 2006.

\bibitem{NilesP01}
I.~Niles and A.~Pease.
\newblock Towards a standard upper ontology.
\newblock In {\em FOIS}, pages 2--9, 2001.

\bibitem{Nowak11}
S.~Nowak, K.~Nagel, and J.~Liebetrau.
\newblock {The CLEF 2011 Photo Annotation and Concept-based Retrieval Tasks}.
\newblock In {\em {CLEF 2011 working notes}}, 2011.

\bibitem{YAGO}
F.~M. Suchanek, G.~Kasneci, and G.~Weikum.
\newblock {YAGO: A Large Ontology from Wikipedia and WordNet}.
\newblock {\em J. Web Sem.}, 6(3):203--217, 2008.

\bibitem{TouschHA12}
A.-M. Tousch, S.~Herbin, and J.-Y. Audibert.
\newblock Semantic hierarchies for image annotation: A survey.
\newblock {\em Pattern Recognition}, 45(1):333--345, 2012.

\bibitem{ZhangIL12}
D.~Zhang, M.~M. Islam, and G.~Lu.
\newblock A review on automatic image annotation techniques.
\newblock {\em Pattern Recognition}, 45(1):346--362, 2012.

\end{thebibliography}

\end{document}